%% file: conference_101719.tex
\documentclass[conference]{IEEEtran}
\IEEEoverridecommandlockouts
\usepackage{cite}
\usepackage{amsmath,amssymb,amsfonts}
\usepackage{algorithmic}
\usepackage{graphicx}
\usepackage{textcomp}
\usepackage{comment}
\usepackage{xcolor}
\usepackage[]{caption}
\usepackage{balance}
\usepackage{tikz}
\usepackage{siunitx}

\usetikzlibrary{3d,calc}

\usepackage{subcaption}
\def\BibTeX{{\rm B\kern-.05em{\sc i\kern-.025em b}\kern-.08em
    T\kern-.1667em\lower.7ex\hbox{E}\kern-.125emX}}
\begin{document}

\title{Programmable EM Sensor Array for Golden-Model Free Run-time Trojan Detection and Localization
}

\author{\IEEEauthorblockN{Hanqiu Wang}
\IEEEauthorblockA{ECE Department, University of Florida \\
Gainesville, FL 32611 USA\\
wanghanqiu@ufl.edu}
\and
\IEEEauthorblockN{Max Panoff}
\IEEEauthorblockA{ECE Department, University of Florida \\
Gainesville, FL 32611 USA\\
m.panoff@ufl.edu}
\and
\IEEEauthorblockN{Zihao Zhan}
\IEEEauthorblockA{ECE Department, University of Florida \\
Gainesville, FL 32611 USA\\
zhan.zihao@ufl.edu}
\and
\IEEEauthorblockN{Shuo Wang}
\IEEEauthorblockA{ECE Department, University of Florida \\
Gainesville, FL 32611 USA\\
shuo.wang@ece.ufl.edu}
\and
\IEEEauthorblockN{Christophe Bobda}
\IEEEauthorblockA{ECE Department, University of Florida \\
Gainesville, FL 32611 USA\\
cbobda@ece.ufl.edu}
\and
\IEEEauthorblockN{Domenic Forte}
\IEEEauthorblockA{ECE Department, University of Florida \\
Gainesville, FL 32611 USA\\
dforte@ece.ufl.edu}
}

\maketitle

\input{content/abstract.tex}

\input{content/Introduction}

\input{content/ThreatModel}

\input{content/PSA}

\input{content/tamper}

\input{content/imple}

\input{content/eval}

\input{content/conclu}

\input{content/ack}

\bibliographystyle{IEEEtran}
\bibliography{bibs/conference_101719}

\balance

\end{document}

%% file: content/abstract.tex
\begin{abstract}
Side-channel analysis has been proven effective at detecting hardware Trojans in integrated circuits (ICs). However, most detection techniques rely on large external probes and antennas for data collection and require a long measurement time to detect Trojans. Such limitations make these techniques impractical for run-time deployment and ineffective in detecting small Trojans with subtle side-channel signatures. To overcome these challenges, we propose a Programmable Sensor Array (PSA) for run-time hardware Trojan detection, localization, and identification. PSA is a tampering-resilient integrated on-chip magnetic field sensor array that can be re-programmed to change the sensors' shape, size, and location. Using PSA, EM side-channel measurement results collected from sensors at different locations on an IC can be analyzed to localize and identify the Trojan. The PSA has better performance than conventional external magnetic probes and state-of-the-art on-chip single-coil magnetic field sensors. We fabricated an AES-128 test chip with four AES Hardware Trojans. They were successfully detected, located, and identified with the proposed on-chip PSA within 10 milliseconds using our proposed cross-domain analysis. 

\end{abstract}


%% file: content/Introduction.tex
\section{Introduction}
\label{sec:intro}
Hardware Trojans (HTs) are malicious modifications inserted into ICs by adversaries during the fabrication or design process.
These alterations can disrupt system functionality, causing security and privacy issues such as privacy leakage, denial-of-service and privilege escalation, etc.
As the reliance on third-party designs and manufacturing continues to grow, the threat posed by HTs becomes more alarming.
Over the years, researchers have discovered that HTs often produce detectable anomalies through side-channels like power consumption and electromagnetic (EM) emanations.
The anomalies in side-channels can be detected after the chip is fabricated, either during the test phase or in run time, but often require costly external sensing setups and a golden model, a Trojan-free circuit used as a reference. This makes deploying some of these approaches in run time not applicable. Moreover, many side-channel oriented techniques suffer from low resolution of the collected traces.
As a result, on-chip sensors have gained attraction as a preferred solution. These sensors can intrinsically collect traces from EM side-channels through magnetic field coupling at run time with higher resolutions, thus ruling out the need for external measurement equipment (e.g., external probes, moving stages, etc.). 
Previous research by Jiaji et al.~\cite{jiaji} used a single coil covering the entire chip area, which has low signal-to-noise ratio (SNR) issues. Consequently, this approach couldn't detect small HTs with few gates and required over 10,000 measurements to detect larger HTs. Its long mean time to detect (MTTD) limits its practical effectiveness.

\begin{figure*}[t]
    \centering
     \begin{subfigure}[t]{0.4\textwidth}
    \centering
    \includegraphics[width=\textwidth]{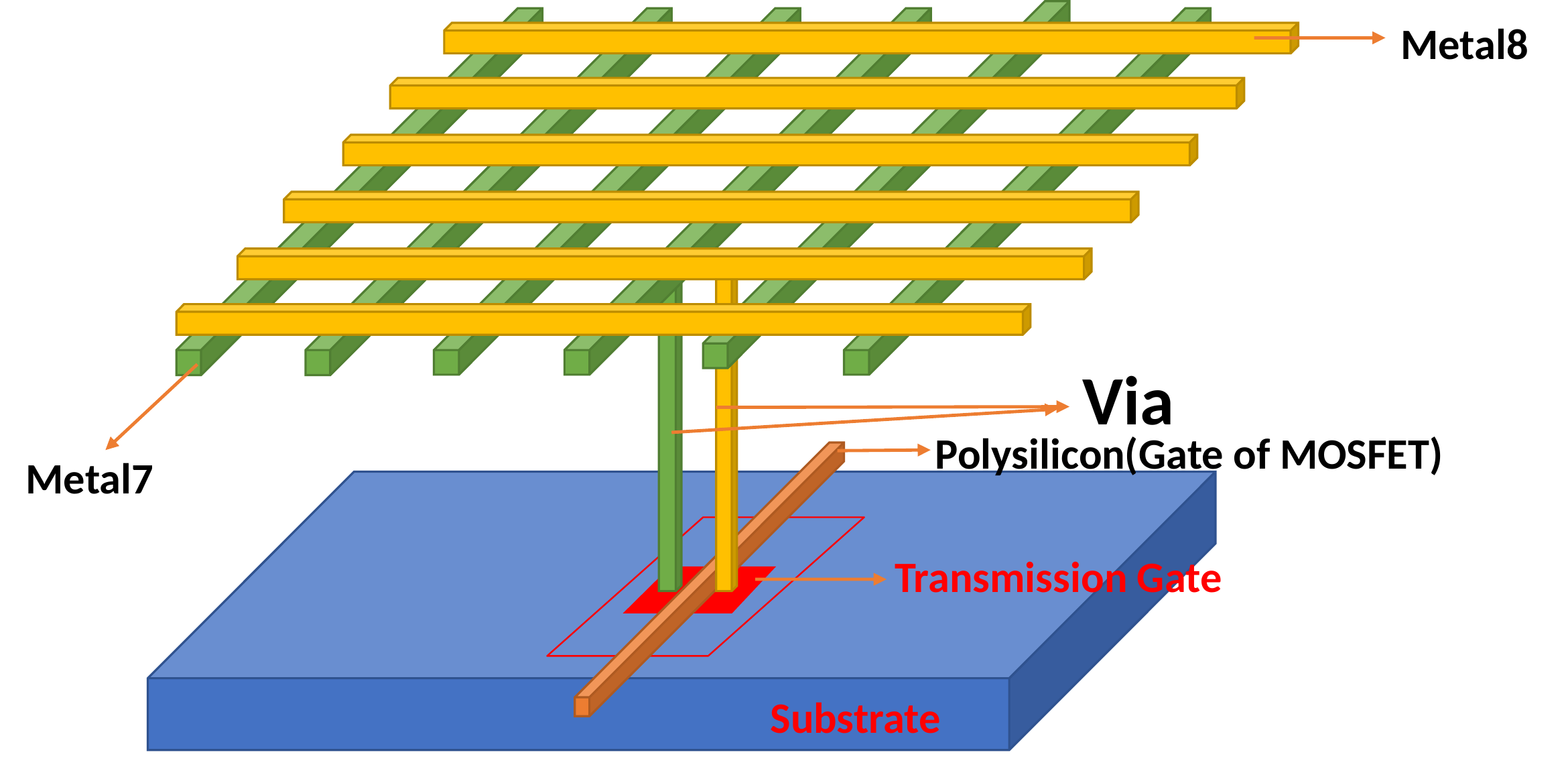}
    \caption{}
    \label{fig:3dpsa}
    \end{subfigure}
    \hspace{-20pt}\begin{subfigure}[t]{0.27\textwidth}
    \centering
    \includegraphics[width=\textwidth]{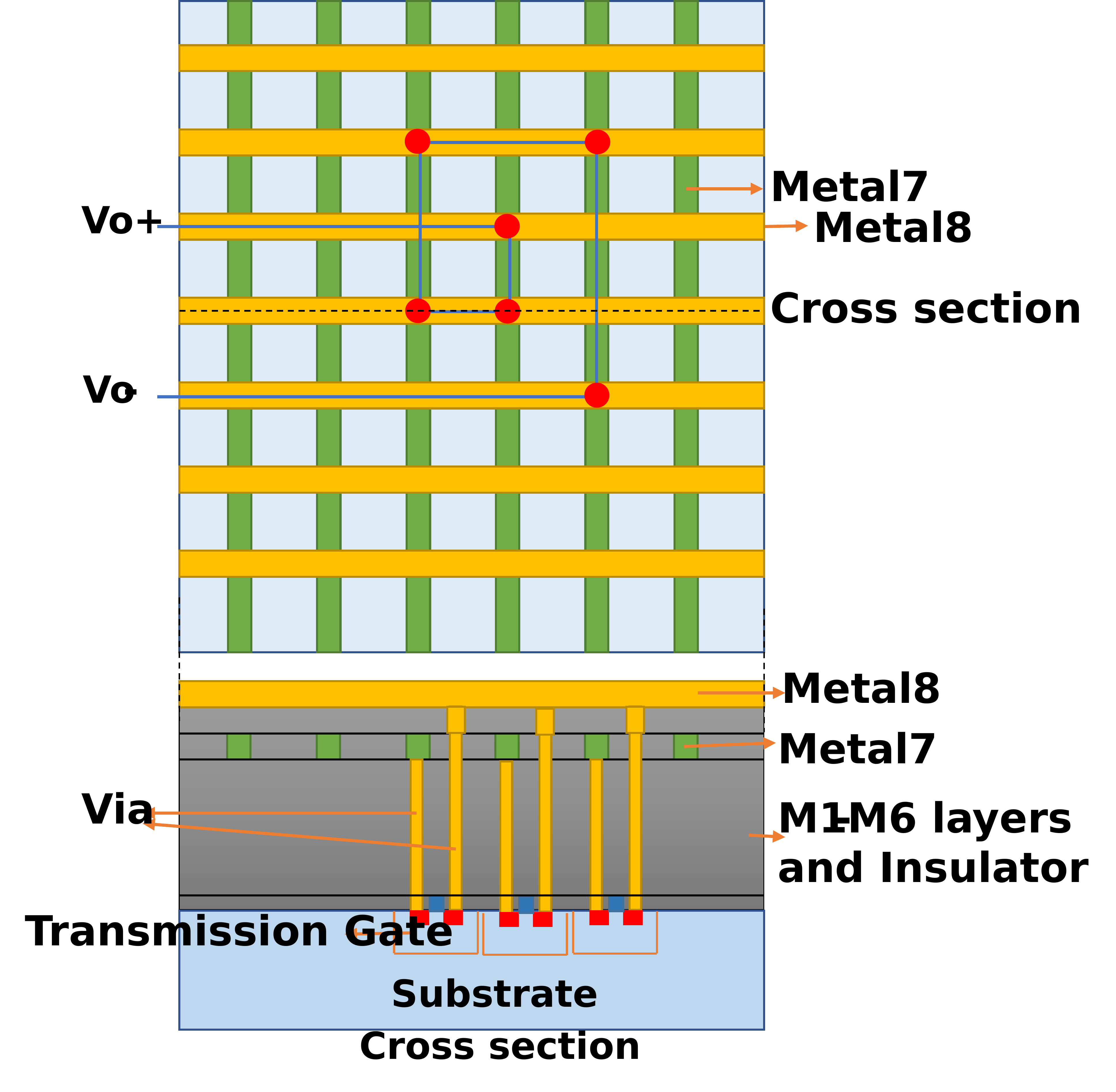}
    \caption{}
    \label{fig:crosssec}
    \end{subfigure}
    \hspace{-20pt}\begin{subfigure}[t]{0.24\textwidth}
    \centering
    \includegraphics[width=\textwidth]{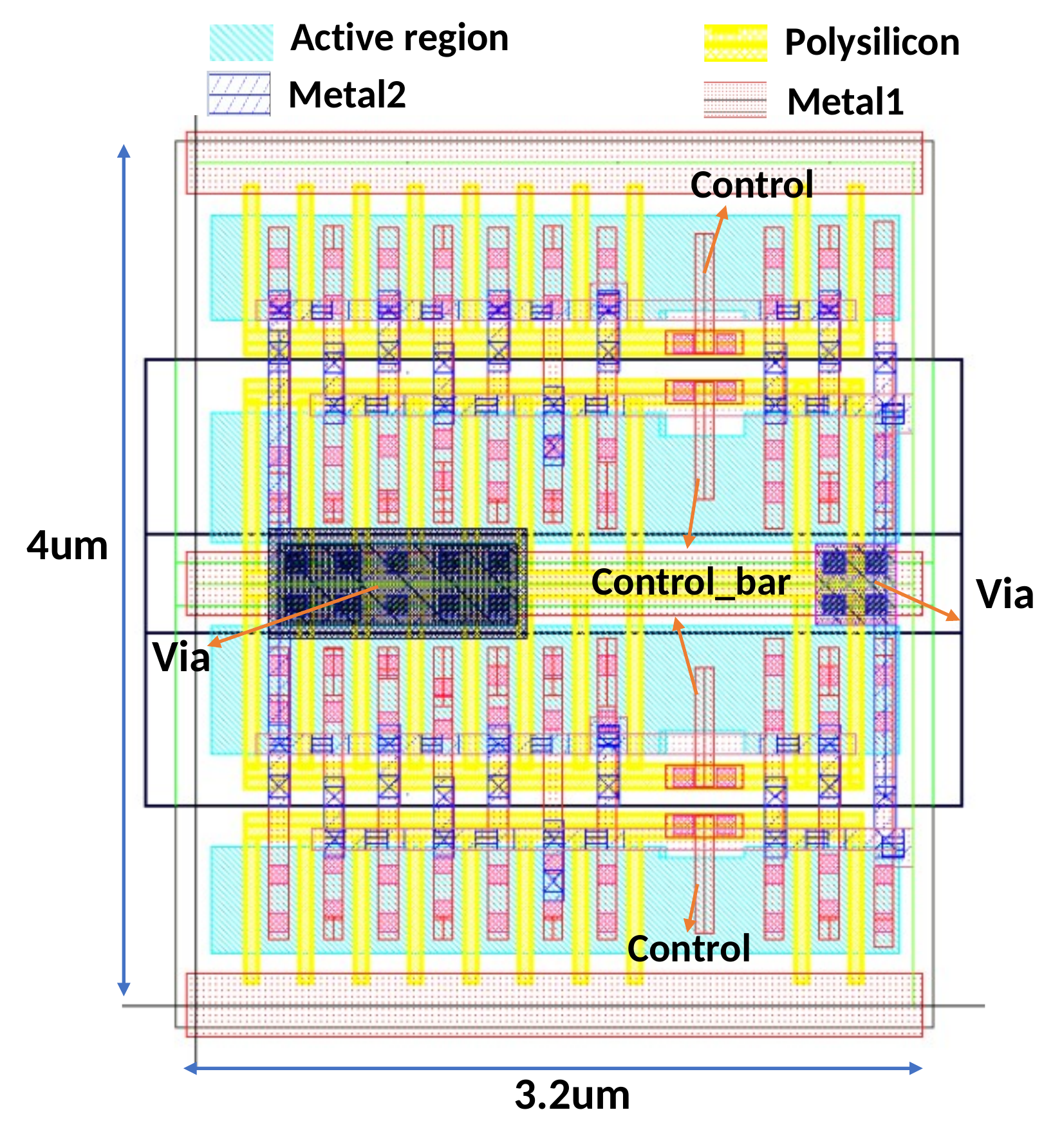}
    \caption{}
    \label{fig:tgate}
    \end{subfigure}

    \caption{(a)3D structure of on-chip PSA where top-level metals form the coils and transmission gate switches in the active layer control the size and shape at each intersection. (Only one MOSFET is shown for simplicity) (b)An example of PSA topology forming a 2-turn coil. The red dots denote the locations where switches are on. (c)T-gate layout.} 
\end{figure*}

In this paper, we propose a Programmable Sensor Array (PSA) to address the limitations of Jiaji's work. Our improved design, shown in Figure \ref{fig:3dpsa}, consists of a crossbar-like structure. The PSA can be programmed to different sizes and shapes, allowing it to cover or uncover specific parts of the chip for better SNR and HT localization. At each crossing point in the wire grid, a transmission gate (T-gate) controls the node's connectivity. This allows the sensor array to have different shapes, locations, and sizes. The size of a single sensor within the PSA can also be programmed to approximately match the size of a HT, ensuring the highest magnetic field emanations from HTs are captured. The improved SNR of PSA enables almost instant runtime detection, reducing MTTD significantly. 
The contributions of this paper can be summarized as follows:
\begin{itemize}
    \item A novel on-chip PSA for EM side-channel measurement with big advantages over existing techniques was proposed, developed and validated.
    \item A silicon demonstration of the proposed PSA to detect and localize active HTs on an AES design implemented with TSMC 65nm technology.
    \item Significant SNR improvements with the proposed PSA compared to prior work.
    \item Experimental validation of the PSA's capability to detect, locate, and identify digital HTs within 10 ms with the proposed novel cross-domain analysis technique.
\end{itemize}


%% file: content/ThreatModel.tex
\section{Threat Model and Prior Work}
\label{sec:back}

\subsection{Run-time Verification Versus Test-phase Verification}

The distinction between run-time and test-phase verifications hinges primarily on the approach to HT detection. During the test phase, efforts are concentrated on the detection of HTs that can be intentionally triggered. The effectiveness of HT detection is evaluated based on the accuracy of feature and trace categorizations as well as the number of attempts needed to activate the Trojan. Most research focuses on developing algorithms to successfully trigger HTs within the minimum amount of time~\cite{chakraborty2009mero}~\cite{huang2016mers}. Conversely, during run time, potential adversaries must first activate an HT before it can be detected if the whole batch of ICs is infected with HT, which is also the assumption of this paper. The primary metric is the duration between the Trojan's activation and its detection. This encompasses recognizing the system's activation and discerning the triggering conditions, often referred to as the Mean Time to Detect (MTTD)~\cite{Forte_temp}.

To reduce the hardware overhead during the evaluations, lightweight measurement and data processing components should be used. Bai et al. developed RASCv2 board, a compact 2cm x 2cm board that can replace an oscilloscope for side-channel analysis to detect malware~\cite{bai2022rascv2}\cite{bai2022real}. The board consists of two analog-to-digital converters (ADCs) for sampling EM and power side-channel traces, an EM antenna for signal capture, and an FPGA for data processing. The data processed by the FPGA are monitored by a security house via an onboard Bluetooth module, making run-time side-channel verification feasible. In this paper, we will integrate the concept of employing the RASC board (manufactured at a trusted facility) for data processing with the PSA's on-chip measurement capabilities to detect HTs in a real-time and trusted manner.

\subsection{EM Side-channel Data Collection Methods}
There are several existing techniques to detect and analyze EM side-channel leakage.
Traditionally, external EM probes and oscilloscopes are used to collect EM side-channel emanation traces over IC packages. The captured signals will be analyzed in either the time or frequency domains to identify HTs \cite{xiaolong, Faezi, Nguyen}. To improve the quality of the collected traces and reduce overhead, He et al. introduced a single-winding on-chip EM sensor covering the whole chip to replace external probes~\cite{jiaji}. In \cite{Nguyen}, Nguyen \textit{et al.} introduced an EM backscattering-based Trojan detection approach. A transmitter antenna injects carrier signals into the IC, and the reflection signals that have been modulated with the activities of HTs are subsequently captured with a receiver antenna to identify HTs. This technique can detect impedance variations resulting from HT activities, even if these HTs have very small current consumption. Compared with our proposed PSA detailed in Section~\ref{sec:meth}, existing measurement techniques have problems of low SNR, limited spatial resolution, complex implementations, etc. These issues result in the limitations of HT detection rates, HT localization, and runtime deployment. A comprehensive comparison will be presented in Section~\ref{subsec:comp}.

%% file: content/PSA.tex
\section{Programmable Sensor Array (PSA)}
\label{sec:meth}
One of the main challenges of on-chip EM sensors is the balance among sensing area, accuracy, and cost. A single large sensing coil covering the entire chip is relatively easy to design and implement; however, it could be inaccurate due to magnetic flux self-cancellation\footnote{Magnetic flux forming a small loop within a large sensor induces no voltage on it.}. Multiple smaller coils on the same layer may improve the overall resolution but would have smaller signal magnitudes and coverage. Using coils across multiplemetal layers can avoid these issues, but with extra cost. To mitigate these problems, we propose the PSA technique that enables the customization of sensor size and position on the chip at run time.

\subsection{PSA Topology and Structure}
PSA utilizes a wire grid that spans two metal layers, with a switch positioned at every intersection. Each switch is essentially a transmission gate (T-gate), depicted in Figure~\ref{fig:tgate}, comprising a single PMOS and NMOS gate connected in parallel. The shape, size, and location of a sensing array can be programmed at run time by controlling specific switches, as shown in Figure~\ref{fig:crosssec}. The size and location of PSA can be programmed via T-gates by connecting or disconnecting different cross-points on the wire grid or lattice. The lattice, comprising several rectangular sensing regions, can be used as part of the sensing array within the PSA.

Adjusting the shape and size of the PSA enables the PSA to circumvent the self-cancellation issue highlighted in \cite{jiaji}, with many advantages. For instance, it facilitates the localization of any detected HTs by reshaping the sensing array. It also eliminates sensing boundary limitations or the need for significant overlap when employing multiple coils.

Implementing such side-channel sensors on chip will raise concerns on PSA's ability to execute side-channel attacks (SCA). Indeed, PSA itself does not protect from SCAs. But if adversaries can physically access the chip and conduct SCAs, it may be vulnerable to SCA regardless of the presence of PSA. Additionally, raw EM traces are not transmitted over any communication channels. Only processed data should be sent, and this can help prevent SCAs. 





\subsection{PSA Comparison with Prior Work}
\label{subsec:comp}
To highlight the improvement of the PSA-based HT detection technique over prior works, we made a comprehensive comparison in Table~\ref{tab:PSA_compar}. This comparison considers five criteria:  HT detection accuracy, spatial resolution of the measurement, required number of traces, SNR, and feasibility for run-time deployment.

The approaches relying on traces collected from conventional external probes~\cite{xiaolong, Faezi} and on-chip single coils~\cite{jiaji} fail to detect small HT with few gates (implemented as T3 in our chip in Table~\ref{tab:trojan_size}). While Nguyen et al.~\cite{Nguyen} reports a 100\% detection rate, their approach shares a common limitation with others: the inability to pinpoint the exact location of the HT on the chip. On the other hand, our PSA approach not only ensures a 100\% detection rate but also stands out with the enhanced spatial resolution, providing the distinct capability of precisely identifying the HTs' physical location.

Due to low SNR, prior works often rely on statistical analysis, requiring numerous measurements to detect Trojans. For instance, Xiaolong et al.~\cite{xiaolong} and Jiaji et al.~\cite{jiaji} compare the Euclidean distance between traces or explore the Euclidean distance distributions. Nguyen et al.~\cite{Nguyen} use Principal Component Analysis and K-means algorithm to categorize the collected spectra. These approaches require 100 to more than 10,000 measurements. 
In contrast, the EM side-channel traces from PSA exhibit sufficiently high SNR with an apparent difference between HT-inactive and HT-active traces. 
Therefore, fewer than ten measurements are needed, leading to a significantly shorter MTTD and faster Trojan detection.

For implementation complexity, unlike prior works necessitating cumbersome measurement devices~\cite{xiaolong,Faezi, Nguyen}, the PSA approach does not need external probes so it has high feasibility for run-time deployment.  Fujimoto et al. also exploited the on-chip power noise measurement(OCM) to execute Correlational Power Analysis with high SNR~\cite{fujimoto1,fujimoto2}, it is also possible to use such OCM to detect HT, but that requires further investigation.

\begin{table}[]
    \captionsetup{font={small, sc}}
    \centering
    \caption{Comparison of EM side-channel data collection methods.}
    \begin{tabular}{|p{2.4cm}|p{1cm}|p{0.8cm}|p{1cm}|p{1cm}|}
        \hline 
         \textbf{Features} &\textbf{External Probe \cite{xiaolong,Faezi}} &\textbf{Nguyen \cite{Nguyen}}
          &\textbf{On-chip Single Coil \cite{jiaji}} 
          & \textbf{PSA (proposed)}  \\
         \hline\hline
         HT Detection rate  & Low & High & Low & High\\
         \hline 
         HT Localization & No & No & No & Yes \\
         \hline
         Measurement\# & $>$10,000 & 100 & $>$10,000 & $<$10  \\
         \hline
         SNR & 14.3dB & N\slash A & 30.5dB & 41.0dB \\
         \hline
         Run-time analysis & No & No & Yes & Yes\\
         \hline
    \end{tabular}
    \vspace{5pt}
    
    \label{tab:PSA_compar}
\end{table}

%% file: content/tamper.tex
\section{PSA's Tamper-resilience}
\label{sec:threat}

Hardware Trojan risk primarily originates from Third Party Intellectual Property (3PIP) vendors and malicious foundries. Our PSA approach is tampering-resilient and can effectively reduce the risks under both attacking scenarios. The effectiveness of PSA in detecting and locating HTs will be proved with the experimental results of Subsection \ref{subsec:HTdetect}. This section introduces a few case studies demonstrating PSA's resistance to tampering from either source.

\subsection{Case 1: Malicious 3PIP}
For malicious 3PIP attacks, as PSA is implemented on the two topmost metal layers after synthesizing the Register Transfer Level (RTL) code and is isolated from the main circuit, it would not interact with any 3PIP. Therefore, any pre-silicon modifications, including all potential 3PIP attacks, made prior to the layout phase will not compromise the PSA.

\subsection{Case 2: Malicious Foundries}
When dealing with malicious foundries, the PSA itself may be modified. However, tampering with such a mixed digital and analog structure is challenging. The adversary must modify additional circuitry, increasing the attack's complexity, thus lowering the attack's success rate. Additionally, any modifications that disable the PSA will trigger alarms during the test phase, as the PSA will return testing values. 

Even if the attacker successfully completes the modifications, designers can easily detect them by reverse-engineering the two topmost metal layers. They are typically the thickest metal layers and are relatively easy to be reverse-engineered. Alternatively, designers can outsource the fabrication of the two topmost metal layers to other trusted foundries, a split-manufacturing method similar to those discussed in~\cite{6513707}. This provides an additional layer of security, making the PSA a robust defense against HT attacks.

%% file: content/imple.tex
\section{PSA implementation on A Test Chip}
\label{sec:imple}

\subsection{Integrating PSA on AES-128 Test Chip}

To assess the performance of PSA in both test and run-time phases we implemented and fabricated the PSA on the two top metal layers of an Advanced Encryption Standard (AES128) encryption design using TSMC65nm technology. The proposed PSA was situated on layers Metal 7 and Metal 8 of the chip, covering the entire chip. It is a lattice including 36 horizontal wires, 36 vertical wires, and 1296 switches. The lattice wire measures 16$\mu$m in length and 1$\mu$m in width. Frequency sweeping is used to determine the optimal length and width that maximize the signal magnitude in the desired frequency range of 10MHz-100MHz.

For this test chip, the entire area was uniformly divided into 16 square sensing areas or sensors. Each sensor shares 33\% of its area with adjacent sensors to ensure adequate circuitry sampling near the borders between any two sensors.

\begin{figure}[t]
    \centering
    \includegraphics[scale=0.23]{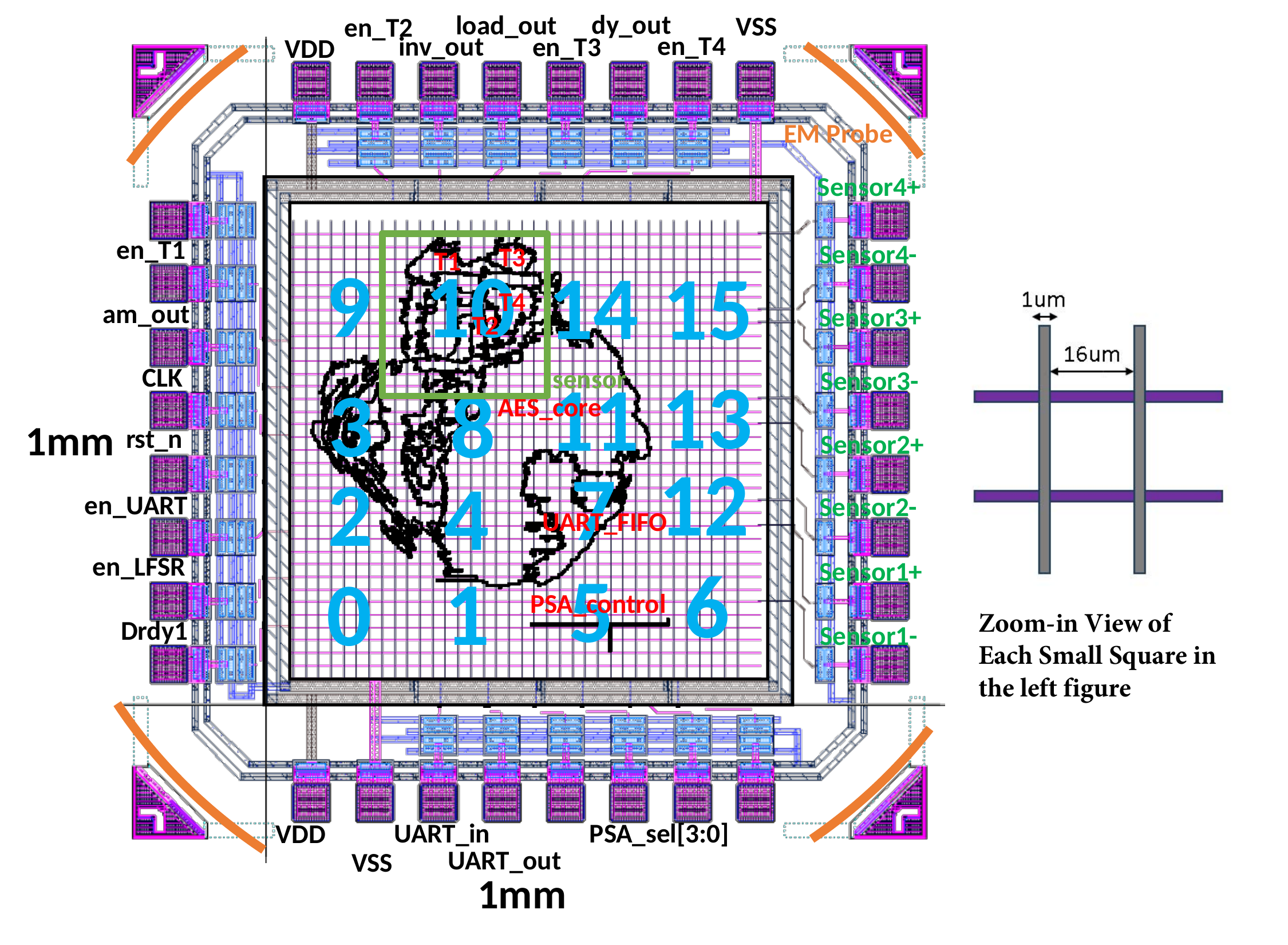}
    \caption{Sensor deployment, IO pin assignment, and Amoeba module view on the AES128 test chip.}
    \label{fig:sensor}
    \vspace{-20pt}
\end{figure}

\textbf{Main Circuit} The underlying main circuit is an AES-128-LUT core, integrated with an RS232 UART communication module \cite{morioka2002optimized}. The clock frequency was set to 33MHz. The layout of the routing wires of the test chip, along with the indices of the PSA sensors, is illustrated in Figure \ref{fig:sensor}.

\textbf{HT Design} The design incorporates four distinct HTs, modified from Trust-Hub \cite{shakya2017benchmarking} \cite{Salmani}. T1 is an amplitude modulation radio carrier Trojan capable of emitting an electromagnetic (EM) wave at a frequency of 750KHz. T2 is a chain of inverters connected to a key wire to amplify its leakage current. If T2 is implanted, attackers could recover the key via power analysis. T3 is a Code Division Multiple Access (CDMA) channel Trojan designed to leak the key, and T4 is a simple denial-of-service Trojan that elevates power consumption, potentially causing the IC to overheat. The four Trojans employ digital standard cells and each has unique triggering conditions. The 4 HT's locations are shown in the Amoeba view in Figure~\ref{fig:sensor}.

\textbf{IO Pin Assignment} The test chip uses a QFN 6mm x 6mm package and has 8 IO pins on each side. The IO pins' names are shown in Figure~\ref{fig:sensor}. The PSA uses the 8 IO pins on the right side. 2 IO pins are needed to measure the induced differential output voltage of one sensor, so we have 4 output channels, from sensor1+/- to sensor4+/-. The 4 sensors on each row use the channel on the same row. For example, sensor 0,1,5,6 in the figure uses sensor1 channel. The PSA control signals use 4 of 8 IO pins on the bottom side. They were decoded into gate signals for T-gates with the fully combinational decoder in the figure.

\textbf{HT Triggering Condition} T1 is activated periodically when a counter reaches 21’h1F\_FFFF under the 33MHz clock. T2 is triggered when the first four bytes of the plaintext are 16’hAAAA. T3 and T4 are the always-on HTs, and we added external enable signals as triggers in experiments. The gate number counts and area ratios of these four HTs are shown in Table \ref{tab:trojan_size}.

\begin{table}[t]
 \captionsetup{font={small, sc}}
 \caption{Trojan gates count and percentage}
 \centering
 \begin{tabular}{|c|ccccc|}
  \hline
  Circuit & Overall & T1 & T2 & T3 & T4 \\
  \hline
  Standard Cell Number & 28806 & 1881 & 2132 & 329 & 2181 \\
  Percentage & 100 & 6.52 & 7.40 & 1.14 & 7.57 \\
  \hline 
 \end{tabular}
 \label{tab:trojan_size}
\end{table}

We assess the performance of the PSA in detecting and locating HTs by activating and deactivating sensors during the execution of AES encryption. As depicted in Figure \ref{fig:sensor}, the green box represents the area of a 6-turn-coil sensor, where most HT circuits are implemented. The orange arcs represent the size of a circle-shape external EM probe. The main circuit primarily falls under sensors 2, 3, 4, 7, 8, 9, 10, 11, and 14. Of these sensors, sensor 10 offers the most coverage of both Trojan payloads and triggers.

\subsection{T-gate Design and PSA Implementation Cost}
Due to the absence of the T-gates that fulfill our requirements in the standard cell library, we designed our own customized T-gates, which are illustrated within a 3.2µm x 4µm customized cell layout in Figure~\ref{fig:tgate}. A single finger of the NMOS and PMOS devices in the design exhibit an aspect ratio of 500nm/60nm and 610nm/60nm respectively to have a similar current conducting capability. For enhanced performance, each of these MOSFETs incorporates 10 fingers. To ensure a low turn-on resistance, the layout features two PMOS and two NMOS devices, effectively creating a pair of T-gates connected in parallel. As a result, this T-gate configuration achieves a resistance of approximately 34 Ohms.


For power overhead, the dynamic power consumption of PSA sensors is negligible and largely contributed by the leakage power from the transistors. Therefore, PSA's impact on the overall power profile is minimal.

For area and resource allocation, the T-gates used in PSA account for an additional 5\% of the total chip area, with the PSA located on metal layers M7 and M8. Despite this, the design effectively preserves routing capacity by aligning PSA wires parallel to main circuit wires, reducing top-layer routing capacity by just 6.25\%.

In contrast to Jiaji's single-coil structure~\cite{jiaji}, which utilizes 100\% of the top layer's routing capacity, PSA proves to be more efficient.

%% file: content/eval.tex
\section{Evaluation on Test Chip}
\label{sec:eval}

\subsection{Test and Data Collection Platform Setup}
    
    

The test chip was mounted on a PCB board with voltage-level shifters and open-loop OP-AMPs. The output of each output channel of the PSA is amplified by a THS4504D OP-AMP with 50dB DC gain and 200MHz UGB, aligning well with our target frequency range from DC to 120MHz.
A 33MHz crystal oscillator is used to generate the clock signal.
The evaluation is conducted while the test chip executes AES-128 encryption.
It receives plaintext from and sends ciphertext to a laptop through serial communications.
During this operation, an oscilloscope or a spectrum analyzer triggered by the rising edge of the clock signal captures the amplified PSA output and generates the related frequency spectrum for further analysis. 
Additionally, we use the zero-span mode of the spectrum analyzer to measure the time-domain signal of the PSA's output at a desired single frequency. 

For run-time deployment, a system integrated on the PCB board can replace the oscilloscopes, spectrum analyzers, and computers to measure and analyze side-channel data.

\subsection{Signal-to-Noise Ratio (SNR) Measurement}
To demonstrate the improved measurement quality offered by the PSA sensor array, we conducted SNR measurements and compared the results with those from an external Langer-EMV LF1 probe.
We employ He's SNR measurement approach as detailed in \cite{jiaji}. Noise traces are first collected from the powered-up chip without any encryption activity. Signal traces are then gathered while the chip performs AES encryption. The SNR is calculated based on the root mean square (RMS) voltage ratio of these two data sets, as demonstrated in Equation \eqref{equation:SNR}.

\begin{equation}
   SNR = 20log \left( \frac{Vrms_{signal}}{Vrms_{noise}} \right)   
   \label{equation:SNR}
\end{equation}

\begin{figure}
    \centering
    \includegraphics[scale=0.25]{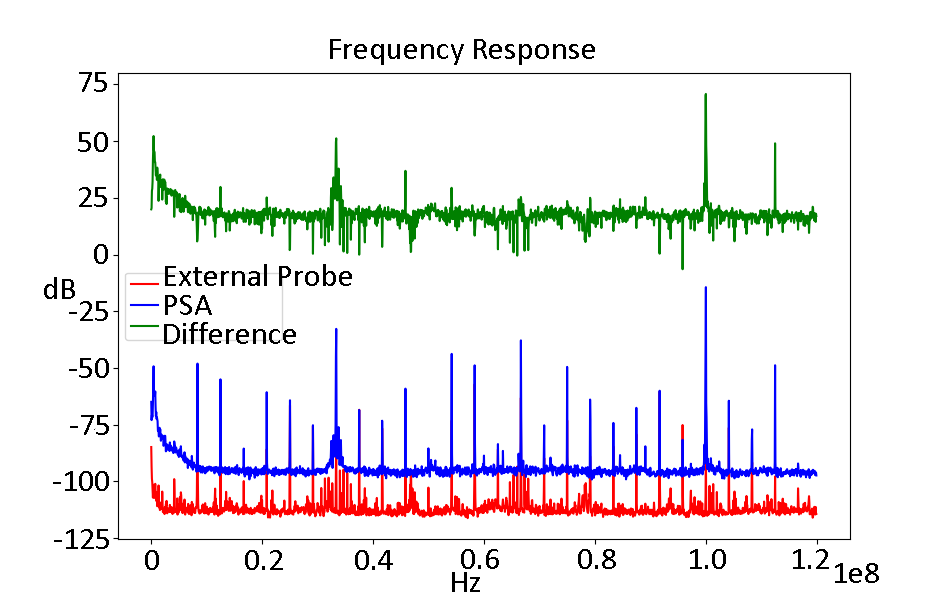}
    \caption{Spectrum magnitude comparison between those from the PSA and an external EM probe.}
    \label{fig:snr}
    \vspace{-20pt}
\end{figure}

The SNR of the PSA is 41.0 dB, surpassing both the external probe's 14.3 dB and the single-coil on-chip sensor's 30.5 dB as reported in \cite{jiaji}.
In Figure \ref{fig:snr}, the green spectrum represents the difference in dB between the spectra from the sensor array and the external probe, demonstrating that the spectrum from the PSA can be up to 55 dB higher than that from an external EM probe. 
Overall, the proposed PSA structure offers significantly superior SNR results in comparison to either external probes or single-coil on-chip sensors.

The best state-of-the-art external probe is the ICR HH100-6 set with a diameter of 100 $\mu$m manufactured by the Langer EMV-Technik. Based on the manufacturers documents, the SNR of ICR probe is approximately 34dB below 120 MHz, so the best probe is still worse than our proposed PSA. 

\subsection{PSA Performance under Different Supply Voltages and Different Ambient Temperatures}

We have conducted comprehensive evaluations of PSA's performance under varying operational conditions to confirm its suitability for runtime deployment across a wide range of applications. We performed simulations and experiments at different supply voltages, spanning from 0.8V to 1.2V, which covers the voltage supply voltage range for TSMC 65nm chips. Additionally, we carried out simulations under different ambient temperatures, ranging from \SI{-40}{\celsius} to \SI{125}{\celsius}.

\subsubsection{Voltage}

Theoretically, a supply voltage increase results in a reduction of the turn-on resistance of T-gates. In the Virtuoso simulation, there was only a 4dB drop in the impedance of a single PSA sensor when the voltage supply was raised from 0.8V to 1.2V. We also made an experiment to measure the current response of the PSA by adding a 70mV frequency sweeping chirp signal to one sensor of the PSA and varying the supply voltage from 0.8V to 1.25V. The current does not change significantly, which agrees with the simulation results.


\subsubsection{Temperature}

Theoretically, an ambient temperature change only slightly affects the impedance of the PSA. Our simulations demonstrate that the impedance remains relatively stable, fluctuating within a range of 4dB. 

\subsection{Runtime Cross-Domain Analysis}
\label{subsec:HTdetect}

\begin{figure*}[htbp]    
    \centering
    \begin{subfigure}[t]{0.18\textwidth}
    \centering
    \includegraphics[width=1.3\textwidth]{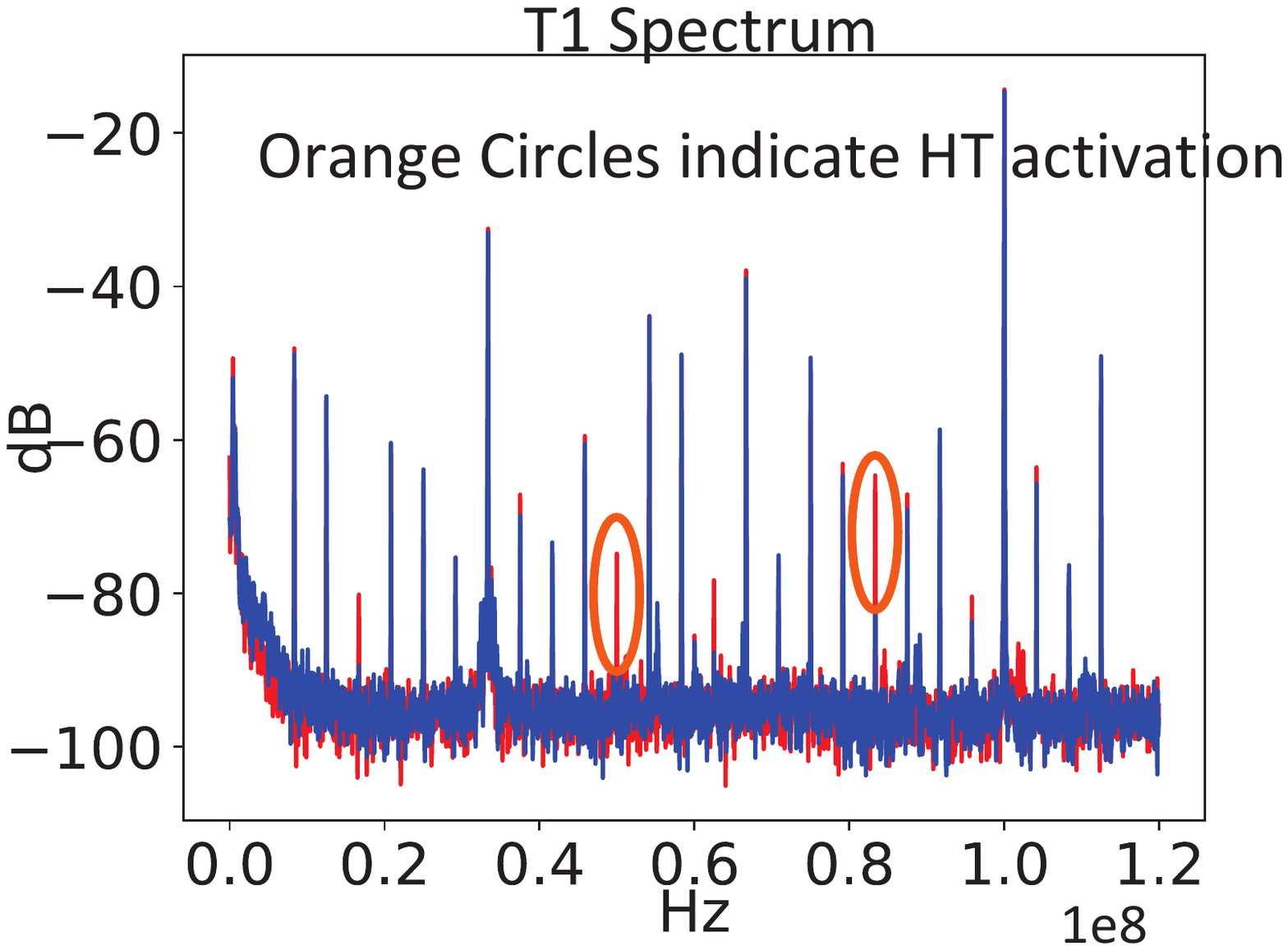}
    \caption{T1 active, Sensor 10}
    \label{fig:T1}
    \end{subfigure}
    \begin{subfigure}[t]{0.18\textwidth}
    \centering
    \includegraphics[width=1.3\textwidth]{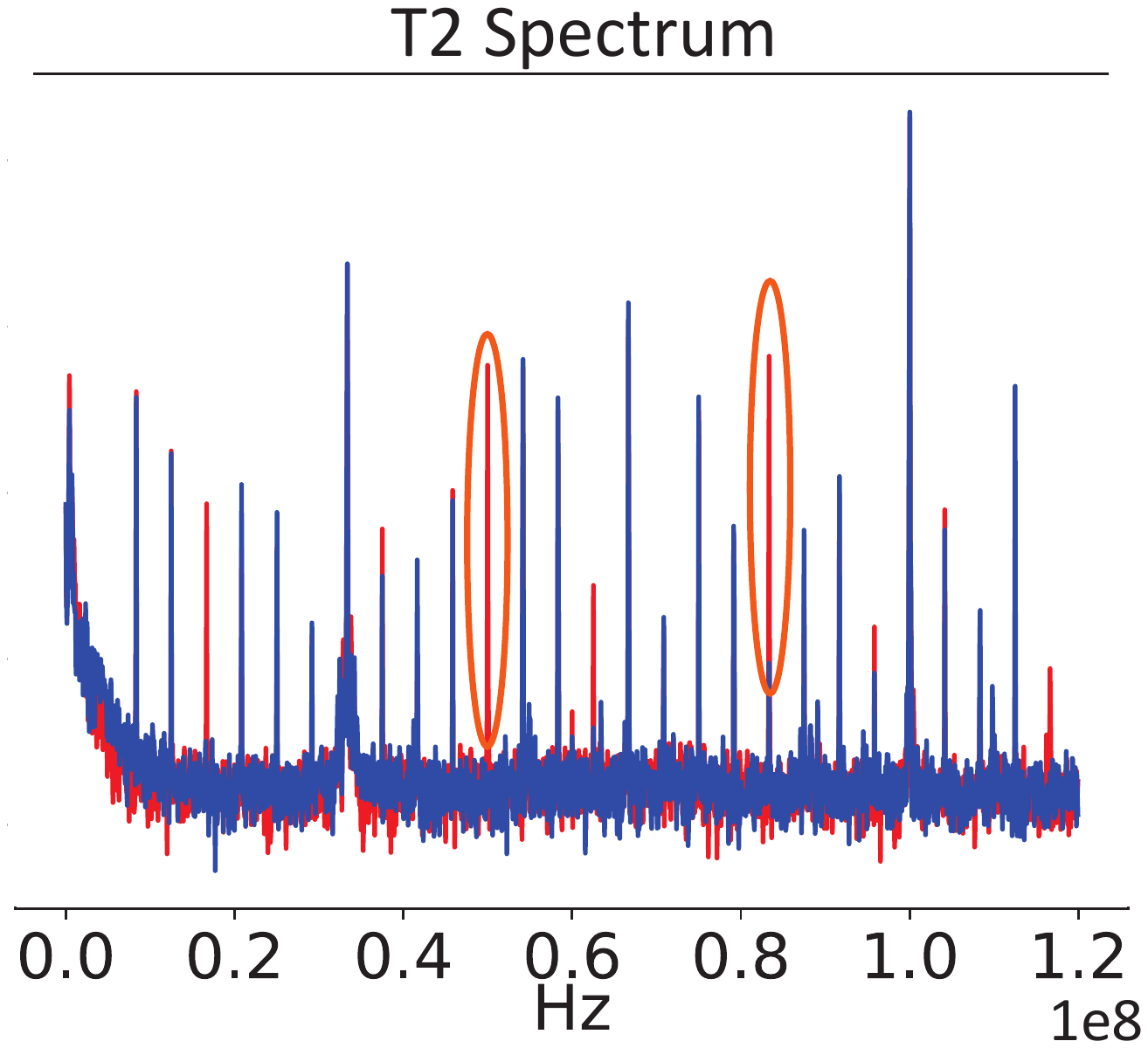}
    \caption{T2 active, Sensor 10}
    \label{fig:T2}
    \end{subfigure}
    \begin{subfigure}[t]{0.18\textwidth}
    \centering
    \includegraphics[width=1.3\textwidth]{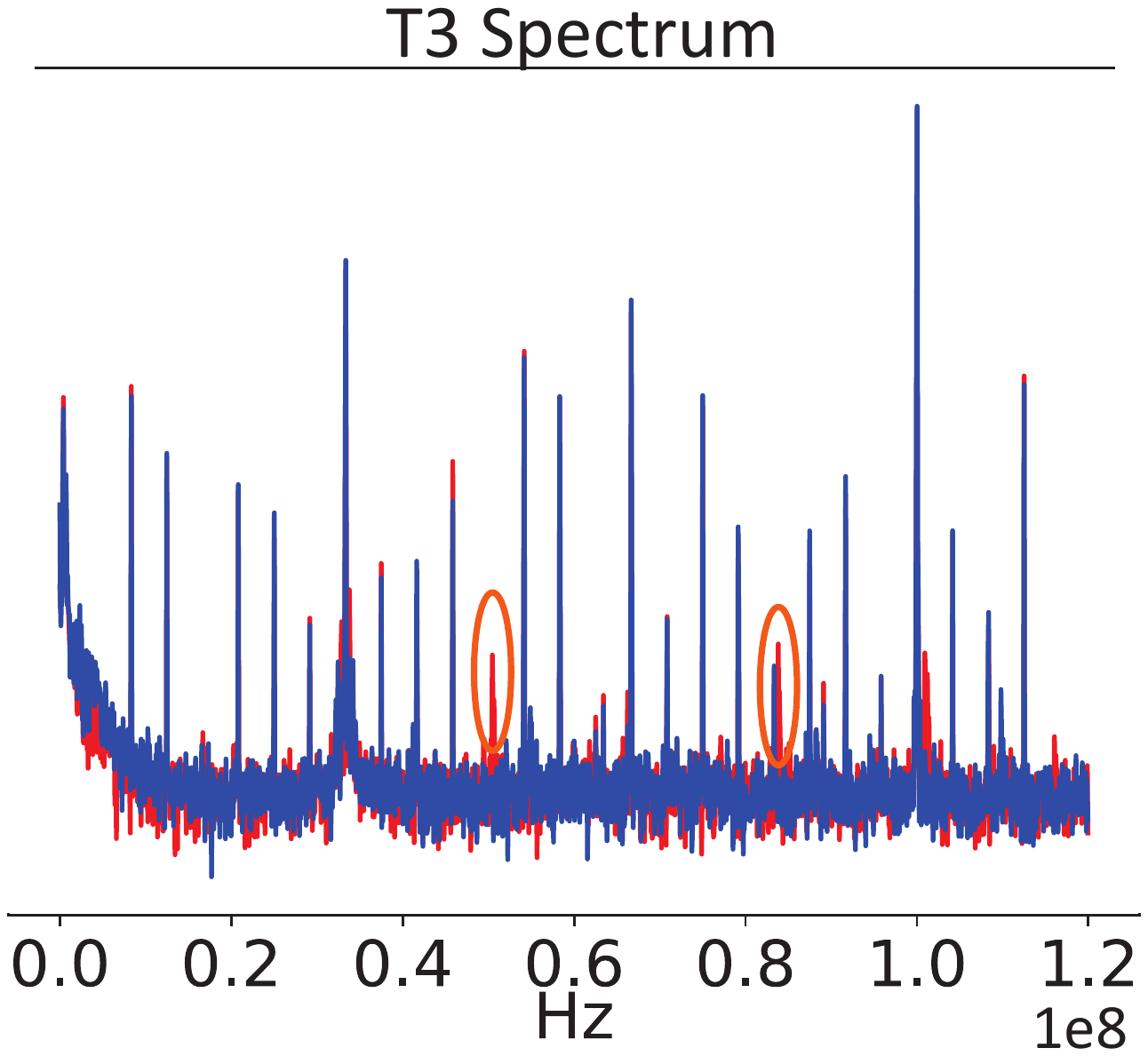}
    \caption{T3 active, Sensor 10}
    \label{fig:T3}
    \end{subfigure}
    \begin{subfigure}[t]{0.18\textwidth}
    \centering
    \includegraphics[width=1.3\textwidth]{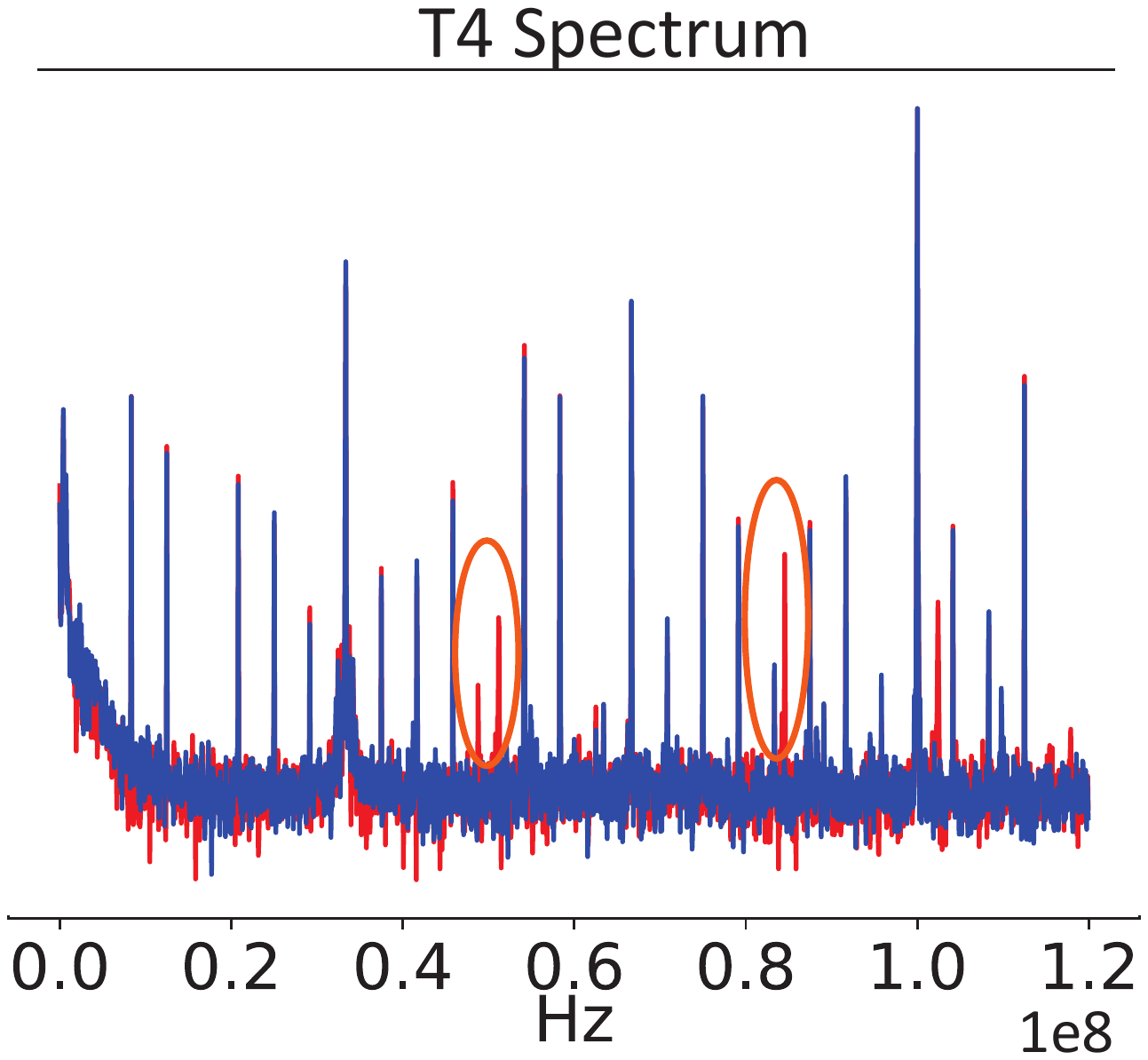}
    \caption{T4 active, Sensor 10}
    \label{fig:T4}
    \end{subfigure}
    \begin{subfigure}[t]{0.18\textwidth}
    \centering
    \includegraphics[width=1.3\textwidth]{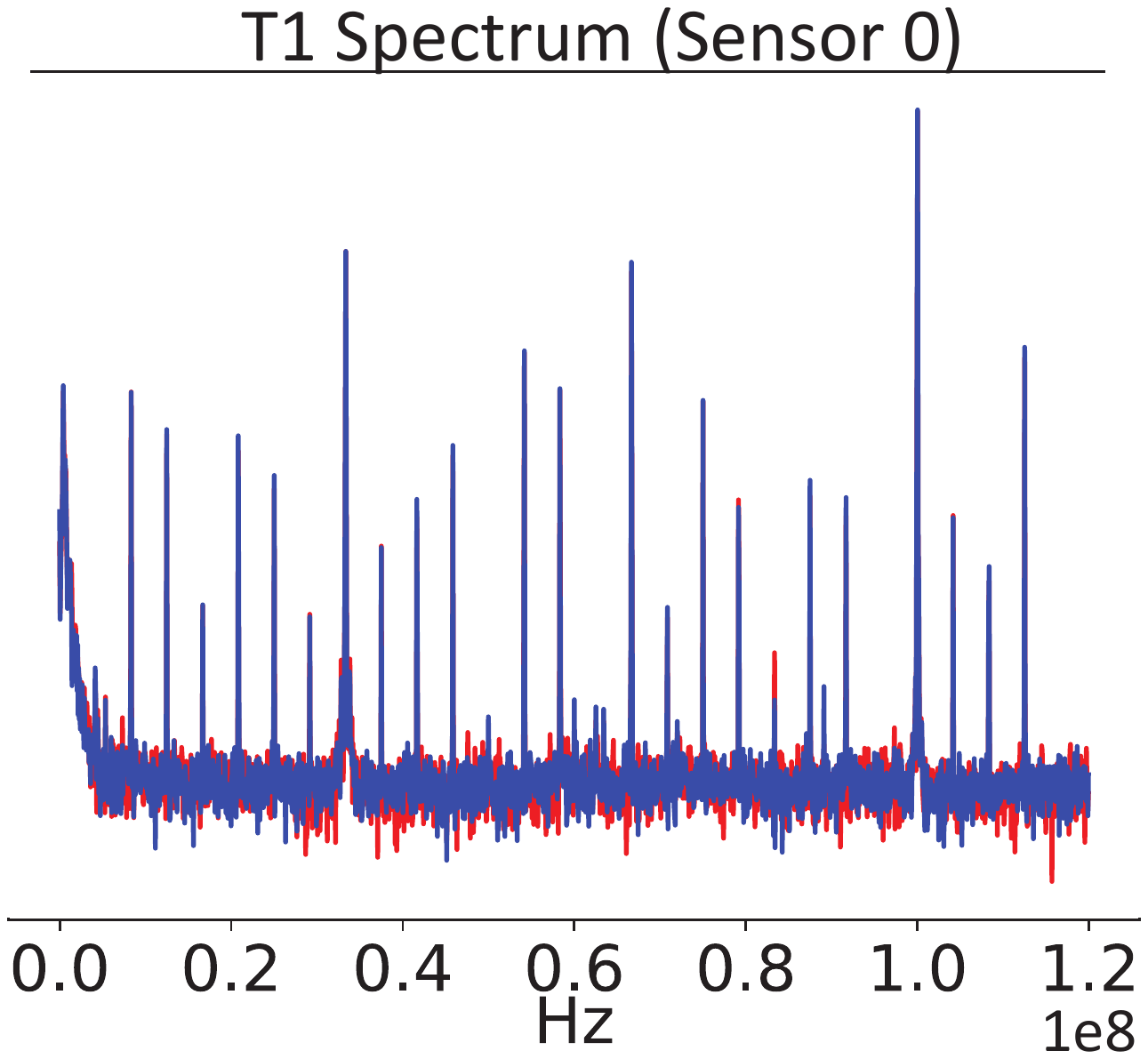}
    \caption{T1 active, Sensor 0}
    \label{fig:0000}
    \end{subfigure}
      \caption{Frequency response captured by sensors 10 and 0 for different HTs: red and blue colors represent the Trojan active and inactive cases, respectively.}
      \label{fig:spec}
      \vspace{-20pt}
\end{figure*}
For each of the 16 sensors, EM traces are recorded under five scenarios: when HTs T1, T2, T3, and T4 are individually activated and in the absence of any active HT. For always-on HTs, we use external enable signals to activate them. Each trace spans a frequency band from DC to 120MHz, populated with 2000 sample points. We averaged five collected traces to derive the spectrum shown in Figure~\ref{fig:spec}. Notably, when examining the output from Sensor 10, as shown in Figure \ref{fig:T1}, \ref{fig:T2}, \ref{fig:T3}, and \ref{fig:T4}, two prominent frequency components at 48MHz and 84MHz, the sideband frequency components of the 1st and 3rd order clock harmonics, show up in the spectrum when HTs are active. We also record the traces from Sensor 0 as shown in Figure~\ref{fig:0000}, where no HTs are implemented beneath the sensor. There is hardly any spectrum difference that can distinguish HT's activities. This result further proves that our proposed PSA has the high spatial resolution needed to locate HTs.

In the experiments, we only need fewer than ten traces collected to detect a HT, resulting in less than 10 ms MTTD, a significant improvement over one single on-chip coil.

\begin{figure}[htbp]
    \centering
     \begin{subfigure}[]{0.24\textwidth}
     \centering
     \includegraphics[scale=0.2]{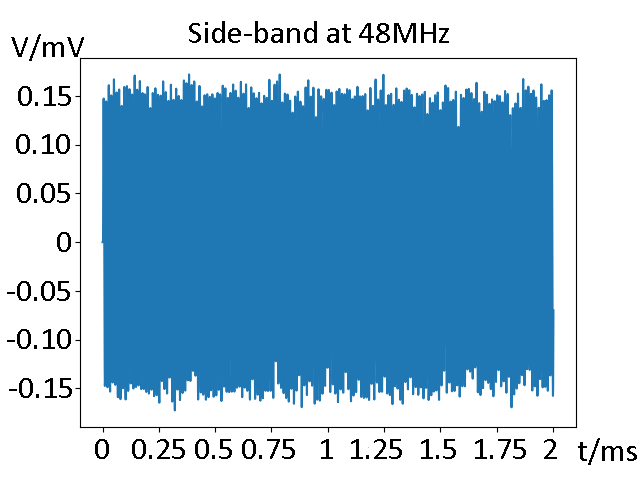}
     \caption{T1 active}
     \label{fig:timeT1}
     \end{subfigure}
     \begin{subfigure}[]{0.24\textwidth}
     \centering
    \includegraphics[scale=0.2]{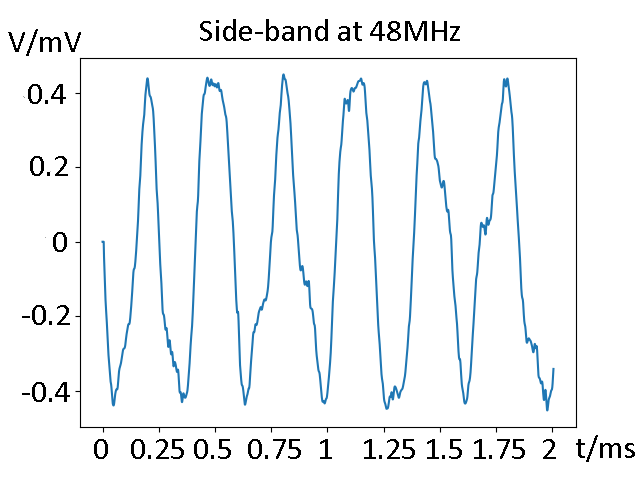}
    \caption{T2 active}
    \label{fig:timeT2}
     \end{subfigure}
    \begin{subfigure}[]{0.24\textwidth}
    \centering
    \includegraphics[scale=0.2]{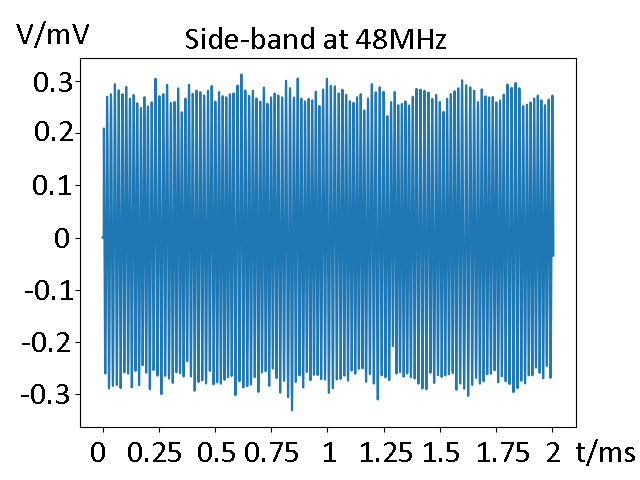}
    \caption{T3 active}
    \label{fig:timeT3}
    \end{subfigure}
    \begin{subfigure}[]{0.24\textwidth}
    \centering
    \includegraphics[scale=0.2]{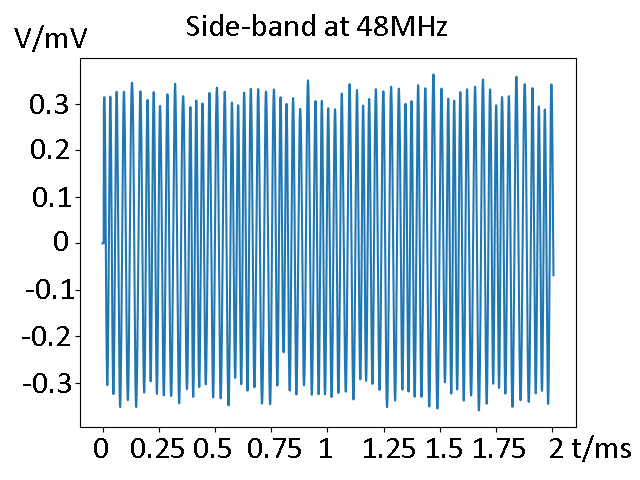}
    \caption{T4 active}
    \label{fig:timeT4}
    \end{subfigure}
      \caption{Time-domain signals of the identified prominent frequency components collected from the PSA sensor 10 were recovered with zero-span mode to differentiate different HTs successfully.}
    \vspace{-10pt}
\end{figure}

After identifying prominent frequency components related to HT's activities, we further switch back to the time domain by employing the zero-span mode of the spectrum analyzer to examine the time-domain waveforms of the two frequency components. Zero-span mode allows us to analyze the pattern of a time-domain signal at a desired fixed frequency. The time-domain waveforms at 48MHz when T1, T2, T3 and T4 are active are shown in Figure \ref{fig:timeT1}, \ref{fig:timeT2}, \ref{fig:timeT3}, and \ref{fig:timeT4}. It is shown that even if different Trojans leaked their information at the same frequency, the difference in their time-domain signals at 48MHz can still clearly differentiate different Trojans. 
This is because different HTs result in different modulation patterns for clock frequency, and the information on modulation patterns is included in the time-domain waveforms of the side band frequency components. 
Therefore, by identifying the unique patterns in the time-domain signals, we can successfully classify all 4 HTs without full supervision.

%% file: content/conclu.tex
\section{Conclusion}
\label{sec:conc}
In this paper, we proposed an integrated on-chip EM sensor array, PSA, that can be programmed to change its shape, size, and location. The PSA was implemented and tested in a 65nm technology node. The proposed PSA has a lower overhead than existing state-of-the-art on-chip coil sensor designs with much higher SNR and the capability of locating HTs. Furthermore, the MTTD during the runtime verification has been significantly reduced to less than 10 ms. The proposed cross-domain analysis can identify prominent frequency components that include the activity information of different Trojans in the frequency domain and differentiate different Trojans by examining the time-domain signals of these prominent frequency components by switching to the time domain. 

%% file: content/ack.tex
\section{Acknoledgement}

This work was supported by the Office of Naval Research under Award Number N00014-19-1-2405.

%% file: conference_101719.bbl
\begin{thebibliography}{10}
\providecommand{\url}[1]{#1}
\csname url@samestyle\endcsname
\providecommand{\newblock}{\relax}
\providecommand{\bibinfo}[2]{#2}
\providecommand{\BIBentrySTDinterwordspacing}{\spaceskip=0pt\relax}
\providecommand{\BIBentryALTinterwordstretchfactor}{4}
\providecommand{\BIBentryALTinterwordspacing}{\spaceskip=\fontdimen2\font plus
\BIBentryALTinterwordstretchfactor\fontdimen3\font minus
  \fontdimen4\font\relax}
\providecommand{\BIBforeignlanguage}[2]{{%
\expandafter\ifx\csname l@#1\endcsname\relax
\typeout{** WARNING: IEEEtran.bst: No hyphenation pattern has been}%
\typeout{** loaded for the language `#1'. Using the pattern for}%
\typeout{** the default language instead.}%
\else
\language=\csname l@#1\endcsname
\fi
#2}}
\providecommand{\BIBdecl}{\relax}
\BIBdecl

\bibitem{jiaji}
J.~He, X.~Guo, H.~Ma, Y.~Liu, Y.~Zhao, and Y.~Jin, ``Runtime trust evaluation
  and hardware trojan detection using on-chip em sensors,'' in \emph{2020 57th
  ACM/IEEE Design Automation Conference (DAC)}, 2020, pp. 1--6.

\bibitem{chakraborty2009mero}
R.~S. Chakraborty, F.~Wolff, S.~Paul, C.~Papachristou, and S.~Bhunia, ``Mero: A
  statistical approach for hardware trojan detection,'' in \emph{International
  Workshop on Cryptographic Hardware and Embedded Systems}.\hskip 1em plus
  0.5em minus 0.4em\relax Springer, 2009, pp. 396--410.

\bibitem{huang2016mers}
Y.~Huang, S.~Bhunia, and P.~Mishra, ``Mers: statistical test generation for
  side-channel analysis based trojan detection,'' in \emph{Proceedings of the
  2016 ACM SIGSAC Conference on Computer and Communications Security}, 2016,
  pp. 130--141.

\bibitem{Forte_temp}
D.~Forte, C.~Bao, and A.~Srivastava, ``Temperature tracking: An innovative
  run-time approach for hardware trojan detection,'' in \emph{2013 IEEE/ACM
  International Conference on Computer-Aided Design (ICCAD)}, 2013, pp.
  532--539.

\bibitem{bai2022rascv2}
Y.~Bai, A.~Stern, J.~Park, M.~Tehranipoor, and D.~Forte, ``Rascv2: Enabling
  remote access to side-channels for mission critical and iot systems,''
  \emph{ACM Transactions on Design Automation of Electronic Systems (TODAES)},
  vol.~27, no.~6, pp. 1--25, 2022.

\bibitem{bai2022real}
Y.~Bai, J.~Park, M.~Tehranipoor, and D.~Forte, ``Real-time instruction-level
  verification of remote iot/cps devices via side channels,'' \emph{Discover
  Internet of Things}, vol.~2, no.~1, p.~1, 2022.

\bibitem{xiaolong}
J.~He, Y.~Zhao, X.~Guo, and Y.~Jin, ``Hardware trojan detection through
  chip-free electromagnetic side-channel statistical analysis,'' \emph{IEEE
  Transactions on Very Large Scale Integration (VLSI) Systems}, vol.~25,
  no.~10, pp. 2939--2948, 2017.

\bibitem{Faezi}
S.~Faezi, R.~Yasaei, and M.~A. Al~Faruque, ``Htnet: Transfer learning for
  golden chip-free hardware trojan detection,'' in \emph{2021 Design,
  Automation \& Test in Europe Conference \& Exhibition (DATE)}, 2021, pp.
  1484--1489.

\bibitem{Nguyen}
L.~N. Nguyen, B.~B. Yilmaz, M.~Prvulovic, and A.~Zajic, ``A novel
  golden-chip-free clustering technique using backscattering side channel for
  hardware trojan detection,'' in \emph{2020 IEEE International Symposium on
  Hardware Oriented Security and Trust (HOST)}, 2020, pp. 1--12.

\bibitem{fujimoto1}
D.~Fujimoto, M.~Nagata, S.~Bhasin, and J.-L. Danger, ``A novel methodology for
  testing hardware security and trust exploiting on-chip power noise
  measurement,'' in \emph{The 20th Asia and South Pacific Design Automation
  Conference}, 2015, pp. 749--754.

\bibitem{fujimoto2}
D.~Fujimoto, N.~Miura, M.~Nagata, Y.~Hayashi, N.~Homma, Y.~Hori, T.~Katashita,
  K.~Sakiyama, T.-H. Le, J.~Bringer, P.~Bazargan-Sabet, and J.-L. Danger,
  ``On-chip power noise measurements of cryptographic vlsi circuits and
  interpretation for side-channel analysis,'' in \emph{2013 International
  Symposium on Electromagnetic Compatibility}, 2013, pp. 405--410.

\bibitem{6513707}
J.~Rajendran, O.~Sinanoglu, and R.~Karri, ``Is split manufacturing secure?'' in
  \emph{2013 Design, Automation \& Test in Europe Conference \& Exhibition
  (DATE)}, 2013, pp. 1259--1264.

\bibitem{morioka2002optimized}
S.~Morioka and A.~Satoh, ``An optimized s-box circuit architecture for low
  power aes design,'' in \emph{International Workshop on Cryptographic Hardware
  and Embedded Systems}.\hskip 1em plus 0.5em minus 0.4em\relax Springer, 2002,
  pp. 172--186.

\bibitem{shakya2017benchmarking}
B.~Shakya, T.~He, H.~Salmani, D.~Forte, S.~Bhunia, and M.~Tehranipoor,
  ``Benchmarking of hardware trojans and maliciously affected circuits,''
  \emph{Journal of Hardware and Systems Security}, vol.~1, no.~1, pp. 85--102,
  2017.

\bibitem{Salmani}
H.~Salmani, M.~Tehranipoor, and R.~Karri, ``On design vulnerability analysis
  and trust benchmarks development,'' in \emph{2013 IEEE 31st International
  Conference on Computer Design (ICCD)}, 2013, pp. 471--474.

\end{thebibliography}
